\documentclass[12pt]{iopart}

\usepackage{iopams}
\usepackage{graphicx}
\usepackage{subfigure}
\usepackage{bbm}

\begin{document}

\title[Quenched Disorder in XY Nematics]{Quenched Disorder and Spin-Glass Correlations in XY Nematics}
\author{L Petridis and E M Terentjev}

\address{Cavendish Laboratory, University of Cambridge, J J
Thomson Avenue, Cambridge, CB3 0HE, U.K.}

\ead{emt1000@cam.ac.uk}

\begin{abstract}
We present a theoretical study of the equilibrium ordering in a
$3D\ XY$ nematic system with quenched random disorder. Within this
model, treated with the Replica trick and Gaussian variational
method, the correlation length is obtained as a function of local
nematic order parameter $Q$ and the effective disorder strength
$\Gamma$. These results, $\xi \sim Q^2 e^{1/Q^2}$ and $\xi \sim
(1/\Gamma)e^{- \Gamma}$, clarify what happens in the limiting
cases of diminishing $Q$ and $\Gamma$, that is near a phase
transition of a pure system. In particular, it is found that
quenched disorder is irrelevant as $Q\to 0$ and hence does not
change the character of the continuous $XY$ nematic-isotropic
phase transition. We discuss how these results compare with
experiments and simulations.
\end{abstract}


J. Phys. A: Math. Gen. $\mathbf{39}\ 9693-9708 \ (2006)$

\section{Introduction \label{intro}}

Classical nematics possess long range orientational order as a
result of anisotropically biased pair interaction between their
long rod-like molecules. Below a certain temperature such systems
are in the ordered phase, where the rods are on average aligned
with their long axes parallel to each other. Macroscopically this
preferred direction is defined by a unit vector called the
director, labelled as $\mathbf{n}$ \cite{degen}. In this paper we
investigate the effects a quenched random field -- disorder
coupled to the local order parameter -- has on the ordering of
$XY$ nematics. This is a special sub-class of systems where the
orientational ordering direction is confined to a plane (hence the
name $XY$), while the overall system dimensionality may remain the
usual $3D$ (or any other). This is of course fully analogous to
the famous $XY$ magnets \cite{kosterlitz,kawamura}, but with the
quadrupolar (nematic) order rather than a dipolar. Physically such
systems are increasingly common in thin planar cells of a liquid
crystal, or in new liquid crystalline elastomers made into thin
free standing films with the director confined in their plane.

The presence of quenched disorder is implicit in many systems, as
different as random ferromagnets \cite{chudnovsky82}, vortex
latices in type-$II$ superconductors \cite{larkin70} and randomly
crosslinked nematic elastomers \cite{fridrikh99}. It arises from
impurities whose distribution in the sample and the constraints
they impose on the local field conformation do not change on the
time scale relevant for the evolution of the main ordering.

The influence of quenched disorder is well understood from a
theoretical point of view. In the limit of weak disorder\footnote{
The empirical criterion of ``weak disorder'' is when the quenched
impurities can be treated as acting on the background of
establishing mean ordering field -- as opposed to the strong
disorder when one cannot define the usual order parameter in the
same way as in a pure system.}, it was first shown by Larkin that
impurities lead to the destruction of long range order
\cite{larkin70}. This result was then generalized by Imry and Ma who
demonstrated that, in less than four dimensions, an arbitrarily weak
random field confines the long range order to hold only at length
scales smaller than the ``domain size" $\xi$. The system breaks into
uncorrelated regions and therefore loses overall order and alignment
\cite{larkin70,imry75}. Much work has been done on $XY$ random
anisotropy magnets, showing magnetization correlations to decay
exponentially \cite{chudnovsky82,chudnovsky86}: $\left\langle
\mathbf{m}(0)\cdot\mathbf{m}(r)\right\rangle \,\sim\,e^{-r/\xi}$.
More sophisticated analysis was done on flux lattices with a two
component order parameter ($N=2$, $d=3$) whose Hamiltonian
incorporates an elastic term and a random potential
\cite{giamarchi95}. It shows that correlations in fact decay as a
power-law $\left(r/\xi\right)^{-1}$, giving rise to quasi-long range
 order (QLRO). A functional renormalization group analysis gives a
similar behavior for $XY$ ferromagnets, but also finds that
quasi-long range order is absent in the case of a magnetization
field with more than three components \cite{feldman00}. Our work
follows a similar line to Ref~\cite{giamarchi95} and also predicts
QLRO. However, another non-pertubative functional renormalization
group analysis disputes the presence of QLRO in $N=2$ $d=3$, where
the system is found to always stay disordered \cite{tissier06}. The
authors find that for $N=2$ the critical dimensionality below which
QLRO exists to be $3.8$.

In many systems with such glassy nature of macroscopic ordering, the
experimental observation of key theoretical predictions is difficult
and often indirect. The loss of the long range orientational order
is directly seen in equilibrium polydomain nematic elastomers. These
materials, formed by crosslinking liquid crystalline polymers in
their isotropic phase, have inherent quenched disorder. There are
two plausible explanations why disorder is present, related to each
other. Residual heterogeneous strains established at the moment of
crosslinking give rise to random stresses (a), acting locally on the
nematic order \cite{uchida00}. Alternatively (b), the locally
anisotropic crosslinks provide randomly oriented anisotropy axes
whose disordering effect competes with Frank elasticity, which
favors a uniform director configuration \cite{fridrikh99}. There is
also a model essentially treating an ad hoc random temperature in a
nematic system \cite{selinger} but its approach, neglecting the
fundamental couplings of nematic ordering to the underlying elastic
network is not relevant for our work. It may be possible that a
careful analysis of director correlations could resolve the physical
origin of quenched disorder due to the network crosslinks, since
computer simulations based on model (a) predict a faster than
exponential decay of director correlations \cite{uchida00}, whereas
simulations based on model (b) find the decay to be exponential
\cite{yu98}. Nevertheless, in both simulations the disorder
correlation length has been a rapidly decreasing function of the
effective disorder strength. The dependance is exponential and
therefore poses the problem that the correlation length does not
diverge at the limit of vanishing disorder.

In practice, the polydomain nematic elastomers have the following
features.  On cooling below the nematic-to-isotropic transition
temperature ($T_{NI}$) a high order parameter phase is formed.
Measurement of the order parameter is difficult, with perhaps only
NMR a suitable technique \cite{disch94,lebar05}. There are
differing reports in the literature over the years, in all cases
finding the transition to be continuous -- approaching a critical
point in better quality samples. It is well established that deep
in the nematic phase the orientation of the nematic director is
very non-uniform: $\mathbf{n}$ varies continuously across the
sample, pointing roughly along one direction across a very small
region of space: following a classical spin-glass pattern which is
perhaps misleadingly called a ``polydomain" structure.
Correlations between director orientations at different points in
space decay rapidly and eventually vanish at distances much larger
than $\xi$ so that long range order is eventually lost
\cite{fridrikh97}. This characteristic length scale $\xi$, is
often called the domain size or correlation length. It is well
established to be of the order of microns \cite{clarke98,elias99}
in many different systems. Therefore, light passing through such a
sample is strongly scattered on birefringent regions with randomly
oriented optical axis. As a result, such a sample is completely
transparent at high temperatures, but becomes opaque below
$T_{NI}$ due to the multiple scattering of the disordered nematic
phase. The dependence of $\xi$ on the magnitude of the order
parameter ($Q$) and the effective strength of quenched disorder
($\Gamma$) is the main focus this work. The experimental work
quoted above was done in thin films leading us to consider
directors that are confined to the $xy$-plane, though being
dependent an all three spatial coordinates. This choice would also
adequately describe nematic membranes \cite{03ximu,06biscari}.

The structure of the paper is as follows. In Section~$2$ we
summarize a physical model of quenched disorder in nematic systems
following \cite{fridrikh97}. Section~$3$ applies the Gaussian
Variational Method \cite{mezard91} to this problem. We find that
the Replica symmetry is broken in this system and most of the
calculations are performed within the Hierarchical Replica
Symmetry Breaking framework. In Section~$4$ we solve the
variational equations and then obtain $\xi$ as a function of $Q$
and $\Gamma$. We also examine how the phase transition is changed
by presenting a predicted  $Q(T)$ plot. Finally, in Section~$6$,
we conclude by discussing our results and comparing them with
experiments.

\section{Model \label{sec:2}}

\subsection{Sources of quenched disorder \label{subsec:2:1}}

\begin{figure} 
\begin{center}
\resizebox{0.3\textwidth}{!}{\includegraphics{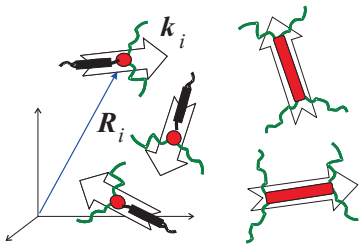}}
\caption{Schematic representation of how crosslinks provide easy
anisotropy axes $\{\mathbf{k} \}$. The nematic director is forced
to be aligned, in the vicinity of the crosslink, with the axes,
which are represent by the arrows. Both the orientation of
$\{\mathbf{k}\}$ as well as the positions of the crosslinks
$\{\mathbf{R}_i\}$ are random. Since the crosslinks are confined
by the network topology, they add quenched disorder to the nematic
system. } \label{crosslinksplot}
\end{center}
\end{figure}

In nematic elastomers crosslinked in the isotropic phase the
network crosslinks act as sources of quenched orientational
disorder. The crosslinks always contain anisotropic groups that
provide easy anisotropy axes $\mathbf{k}$, similar to the sources
in random anisotropy magnets \cite{chudnovsky82}. We assume it is
favorable for the local director to align along $\mathbf{k}$ in
the vicinity of the crosslink. The orientation of the anisotropy
axes as well as the spacial distribution of the crosslinks are
quenched variables. They remain unaltered with time even if
external conditions, such as temperature, change.

We follow previous work \cite{fridrikh97,cleaver} modeling the local
coupling between the nematic order and the random field. For a
crosslink positioned at $\mathbf{R}_i$ with anisotropy axis
$\mathbf{k}_i$ a free energy \ $
 F_i\,=\,-{\gamma} \, \mathbf{k_i}\cdot
\underline{\underline{\mathbf{Q}}}\cdot\mathbf{k_i}$ \ is attributed
to it, where ${\gamma}$ is the coupling strength and
$Q_{ij}\,=\,Q\left(n_in_j\,-\, \frac{1}{2}\,\delta_{ij}\right)$ is
the $XY$ nematic tensor order parameter at spatial position
$\mathbf{R_i}$. The magnitude $Q$ of the nematic order parameter is
different from the Edwards-Anderson order parameter of spin glasses,
but serves the purpose: $Q=1$ refers to a completely ordered state
whereas $Q=0$ to a disordered one \cite{degennes95}. Employing a
continuum density of impurities $\rho\left(\bf r
\right)\,=\,\sum_{\mathbf R_i}\, \delta (\mathbf{r-R_i})$ and
substituting the full expression of $Q_{ij}$ we get the
coarse-grained form of random field energy for the whole sample:
\begin{eqnarray}
F_{r.f.}= - \gamma\,\sum_i \mathbf{k}_i\cdot
\underline{\underline{\mathbf{Q}}}\cdot \mathbf{k}_i\, =
\,-\,\int{d^3\mathbf{r}\, \gamma\,Q \,\rho (\mathbf r) \left(
\mathbf{k \cdot n}\right)^2}
\,+\,\frac{1}{2}\,N_{\rm{cr}}\,\gamma\,Q,
 \label{rfenergy}
\end{eqnarray}
where $N_{\rm{cr}}$ is the total number of impurities
(crosslinks). The positive linear term ($N_{\rm{cr}}\,\gamma\,Q$)
is a byproduct of the requirement that $Q_{ij}$ is traceless. It
cancels out when random disorder is treated properly in
Subsection~\ref{replicasection}, and it is dropped for now.

When the (one-constant approximation) Frank elasticity term is
included, the full continuum Hamiltonian of the nematic system
reads:
\begin{eqnarray}
H\,=\,\int d^3\mathbf{r}\left[ \frac{1}{2}K\left( \mathbf{\nabla}
\mathbf{n}\right)^2\,-\,\gamma\,Q \,\rho (\bf r) \left( \bf{k
\cdot n}\right)^2\right]\ . \label{hamiltonian}
\end{eqnarray}
A simple dimensional argument gives $K\,\sim\, {k_BT}/{a}$, where
$a$ is the interatomic distance below which the continuum limit of
elasticity no longer holds (essentially, the nematic coherence
length \cite{degen}). Both the microscopic and the
phenomenological Landau-DeGennes theories of nematic transition
give the elastic constant $K$ to scale as $Q^2$ for $Q\ll 1$.
Combining these two estimates, we take that for small $Q$ the
elastic constant is approximated by $K\simeq {k_BT}Q^2/{a}$.
Finally, it is noted that the local magnitude of the order
parameter is taken to be constant across the sample and $\nabla
Q\,=\,0$, that is, the effect of quenched disorder lies in the
resulting equilibrium director texture. It is assumed that the
sample is chemically homogeneous and boundary effects are
neglected.

This Hamiltonian has been shown \cite{fridrikh99} to explain
domain formation in nematic elastomers using the classic Imry-Ma
argument \cite{imry75}. Even weak random field destroys long range
order in length scales greater than the correlation length
$\xi_d\sim {K^2}/{\rho_0(\gamma Q)^2}$, where in our notation
$\rho_0$ is the number of crosslinks per unit volume. It is a well
known fact that the correlation length of the order parameter
fluctuations should diverge close to the phase transition.  In
direct contrast, when the Imry-Ma argument is applied to nematic
elastomers the disorder correlation length tends to zero as the
order parameter vanishes ($\xi_d \propto\, Q^2$) because $K$ is a
quadratic functions of $Q$. In Section~\ref{resultssection} we are
able to clarify this rare shortcoming of the Imry-Ma argument.

\subsection{Replica method \label{replicasection}}

We are interested in results averaged over the random
distributions of the quenched variables $\rho(\bf r)$ and $\bf k$
because these variables cannot be controlled experimentally. In
other words, one looks for the macroscopic averages over a variety
of realizations of $\rho(\bf r)$ and $\bf k$. Crosslinking the
sample above $T_{NI}$ makes the easy anisotropy axes point at
random directions in the $XY$ plane, with an isotropic probability
of orientation $P({\bf k})\,=\,\frac{1}{2\pi}$. The crosslinks are
dispersed randomly in the sample with the density
$\rho(\mathbf{r})$ and the probability that a particular
distribution $\rho$ occurs Gaussian \cite{edwards88}:
\begin{eqnarray}
 P[\rho]\,\sim\,\exp\left[- \int{d^3r
\frac{(\Delta\rho)^2}{2\rho_0}}\right], \label{rhodistribution}
\end{eqnarray}
where $\Delta \rho\,=\,\rho-\rho_0$ is the deviation of local
density from $\rho_0$, the mean density of crosslinks.

Averaging the logarithm of the partition function $Z$ to obtain
the free energy is not algebraically possible, so the Replica
Trick \cite{edwards75} is employed to overcome this difficulty.
The expression for the free energy arising from disorder then
reads:
\begin{eqnarray}
F_{d}\,&=&\,-k_BT\,\langle \log Z \rangle _{\rho, \bf k}\,
=\,\left. { - k_B T\frac{\partial }{\partial m}\left\langle {Z^m }
\right\rangle} \right|_{m = 0} \nonumber \\
 &=&\,-k_BT\,\left. \frac{\partial}{\partial
m}\right|_{m=0} \,\prod_a^m \int\emph{D}\mathbf{n}_a\,\exp
\left[-\beta H_{rep} \right], \label{replicatrick}
\end{eqnarray}
where we now have $m$ identical ``replicas'' of the system,
labelled by the index $a$.  Equation~(\ref{replicatrick}) provides
the definition of the Replica Hamiltonian, which no longer depends
on $\mathbf{k}$ and $\rho$. After averaging over disorder this
Hamiltonian is found to be:
\begin{eqnarray}
H_{rep} \equiv\sum_{a,b=1}^m \int d^3 \mathbf{r} \left\{
\frac{1}{2}K\left( \nabla \mathbf{n}_a\right)^2\delta_{ab}-\Gamma
\left(\mathbf{n}_a \cdot \mathbf{n}_b \right)^2 \right\},
\label{replicahamiltonian}
\end{eqnarray}
where subscribes $a$ and $b$ are the replica indexes and $m$ is
the number of replicas that is set to zero at the end of the
calculation. Parameter $\Gamma$, arising from completing the
Gaussian square with the random-field term in
eq.~(\ref{hamiltonian}), reflects the strength of the disorder and
has a quadratic dependance on the order parameter magnitude:
\begin{eqnarray}
\Gamma \,=\,\frac{\gamma^2\rho_0 Q^2} {8\,k_BT}. \label{Gamma}
\end{eqnarray}
It is noted that all replicas are assumed to have equal disorder
strength and equal magnitude of the order parameter; i.e.
$\gamma_a=\gamma_b$ and $Q_a=Q_b$ for all $ a$ and $b$.

The simplest way to ensure that the director is a unit vector
confined in a plane is to parameterize it by a single angle
$\theta$:
$\mathbf{n}\,=\,\left\{\cos\theta(r)\,,\,\sin\theta(r)\right\}$.
In this notation the Replica Hamiltonian, now a functional of
$\theta$, reads:
\begin{eqnarray}
H_{rep}\left[\theta(r)\right]= \sum_{a,b}^m \int d^3 \mathbf{r}
\left[ \frac{K}{2}\left( \nabla \theta_a\right)^2\,\delta_{ab}\ -\
\frac{\Gamma}{2} \cos 2(\theta_a-\theta_b) \right] , \label{Hrep
2d}
\end{eqnarray}
where the irrelevant constant term $-\Gamma V/2$ arising from the
conversion of $\cos^2\theta$ to $\cos 2\theta$ is dropped.

\section{Gaussian Variational Method\label{GVMsection}}

The cosine in the random field term of Eq.~(\ref{Hrep 2d}) posses
an obstacle to the development of the model. The correlation
function and the free energy can be obtained from
Eq.~(\ref{replicatrick}) as long as the Replica Hamiltonian is
quadratic in $\theta$. This is clearly not the case in our model
and to overcome this obstacle we employ the Gaussian Variational
Method (GVM), first used in the context of random manifolds by
Mezard and Parisi \cite{mezard91}.

\subsection{GVM in Nematic Systems}

A model Hamiltonian $H_0$ is assumed to describe the system,
meaning that all its physical properties, such as correlations and
free energy, are fully defined by $H_0[\theta]$ instead of the
original $H_{rep}[\theta]$. It is essential that $H_0$ is
quadratic in the fields $\theta(\mathbf{q})$, in Fourier space:
\begin{eqnarray}
\beta H_0\,=\,\frac{1}{2}\,\sum_{a,b}^m\int\frac{d^3q}{(2\pi)^3}
\,\theta_a(q)\, G_{ab}^{-1}(q)\,\theta_b (q). \label{H0}
\end{eqnarray}
Here and throughout the paper we use the shorthand notation
$\beta=1/k_BT$ that makes all energy functions non-dimensional.
The (unknown) propagator can be written as:
\begin{eqnarray}
G_{ab}^{-1}\,=\,\beta K\,q^2\,\delta_{ab}\,-\,\sigma_{ab}\,,
 \label{G^-1 2d}
\end{eqnarray}
where $K\,q^2\,\delta_{ab}$ is chosen to match the elastic
contribution in $H_{rep}$, and $\sigma_{ab}$ is a set of yet
undetermined parameters approximating the random field effects.
The true free energy of the system can be cast as an expansion in
$\langle H_{rep}-H_0\rangle $, assumed small:
\begin{eqnarray}
F \,\approx\,F_0 + \langle H_{rep}-H_0 \rangle_{H_0} ,
\label{Fvar1}
\end{eqnarray}
where $F_0= -k_BT\, ({\rm det} \,G)^{1/2}$ and the angular
brackets indicate averaging with the Gaussian $e^{-\beta H_0}$ as
measure. The ultimate aim is to minimize $F[G]$ with respect to
variations in $G_{ab}$ and obtain the best upper bound for $F$.
Using Eq.~(\ref{H0}), the average of Eq.~(\ref{Fvar1}) leaves the
variational energy density:
\begin{eqnarray}
\fl \beta F_{var} = \frac{1}{2} \int_q \left[ -{\rm tr}
\log\mathbf{G} + \sum_a^m \beta K q^2 \, G_{aa} \right]  -
\frac{1}{2} \sum_{a, b}^m \beta \Gamma  \exp\left[-2 \int_q
\left(G_{aa}-2G_{ab}+G_{bb}\right) \right]\,, \nonumber
\end{eqnarray}
where notation $\int_q$ is a shorthand for the full  $\int
{d^3q}/{(2\pi)^3}$ and the irrelevant constant $\langle H_0
\rangle=-k_BT/2$ has been dropped. Before proceeding with the
minimization it is convenient to define a new function that
appears regularly in the random field term:
\begin{eqnarray}
B_{ab}\,=\,\int_q \left(G_{aa}-2G_{ab}+G_{bb}\right) \ .
\label{Bab}
\end{eqnarray}
It is related to the correlations between replicas since
$\langle[\theta_a(r)\,-\,\theta_b(r)]^2\rangle\,=\,B_{ab}$.

Demanding $F_{var}$ to be stationary with respect to variations on
$G_{ab}$ provides a relation that determines $\sigma_{ab}$.
\begin{eqnarray}
\frac{\delta\,\beta F_{var}}{\delta\,G_{cd}} = 0\ \ \Rightarrow  \
G_{ab}^{-1}=\beta K q^2\,\delta_{ab}-4\beta \Gamma  \left(
e^{-2\,B_{ab}}-\delta_{ab}\sum_c\,e^{-2\,B_{ac}}\right)
\label{1diff}
\end{eqnarray}
Comparing this expression with $F_{var}$, it is easily seen that
the stationary equation for the matrix of parameters
$\mathbf{\sigma}$ takes the form:
\begin{eqnarray}
\sigma_{ab}\,=\,{4 \beta \Gamma}\,\left[
e^{-2B_{ab}}\,-\,\delta_{ab}\,\sum_c\,e^{-2B_{ac}}\right],
\label{sigma general}
\end{eqnarray}
The above equation has a reassuring property, that $\sigma$ is
zero when there is no disorder in the system  ($\Gamma=0$). After
all, $\sigma_{ab}$ was introduced to approximate the random field
imposed by the crosslinks.

It is convenient to decompose the equation into two separate
conditions \cite{mezard91}; one for the off-diagonal and the
separate for the purely diagonal part of $\sigma_{ab}$. The
off-diagonal part is easily obtained by setting the Kronecker
delta to zero in Eq.~(\ref{sigma general}):
\begin{eqnarray}
\sigma_{a\neq b}\,=\, {4 \beta \Gamma V} \,e^{-2B_{ab}} .
 \label{sigma}
\end{eqnarray}
This is a transcendental equation because $B_{ab}$ is itself a
function of $G_{ab}$ and therefore of $\sigma_{ab}$ as well. To be
able solve it a certain form of $\sigma_{ab}$ must be assumed so
that $G_{ab}^{-1}$ can be inverted to obtain $B_{ab}$. The
diagonal part is fixed by noting that summation over the free
index $a$ in Eq.~(\ref{sigma general}) is zero because
$B_{ab}=B_{ba}$. Hence after finding the off-diagonal elements the
condition
\begin{eqnarray}
\sum_{a=1}^m\sigma_{ab}\,=\,0\ .
 \label{sigma_aa}
\end{eqnarray}
provides a way to determine the diagonal matrix element
$\sigma_{aa}$.

\subsection{Replica symmetry \label{RSsection}}

In a first attempt to determine the variational parameter
$\sigma_{ab}$ from the stationary equation, let us assume all the
non-diagonal elements to be equal to $\sigma_{a\neq b}\,=\,
\sigma$, a constant which will be determined by Eq.~(\ref{sigma}).
Using Eq.~(\ref{sigma_aa}) we then find the diagonal part to be:
$\sigma_{aa}\,=\,(1-m)\,\sigma\ $. This scheme based on the
constant-$\sigma$ assumption is called the Replica symmetry (RS)
limit. Under this assumption, the model Hamiltonian acquires the
simple form:
\begin{eqnarray}
G^{-1}_{ab}(q)\,&=&\,\left(\bar{K}\,q^2+m \sigma \right)
\delta_{ab} \,- \, \sigma\,\mathbbm{1}_{ab} \,,\label{G^-1RS}
\end{eqnarray}
where $\bar{K}\,=\beta K$ and $\mathbbm{1}_{ab}$ is the matrix
with all its elements equal to one. The matrix can be trivially
inverted:
\begin{eqnarray}
G_{ab}\,=\,\frac{1}{\bar
{K}\, q^2}\,\delta_{ab}\,+\,\frac{\sigma}
{\bar{K}^2\,q^4}\,\mathbbm{1}_{ab} \label{GRS}
\end{eqnarray}
The stationary equation (\ref{sigma}) is easily solved because
$B_{ab}$ does not depend on $\mathbf{\sigma}$. This is a unique
feature of RS in this problem. We shall see later that less
symmetric forms of $\sigma_{ab}$ generate a transcendental
equation more difficult to deal with. The replica symmetric
solution is:
\begin{eqnarray}
\sigma = {4 \beta \Gamma}\,\exp\left(-4\,\int_q\,
\frac{1}{\bar{K}\,q^2}\right)\,= \,{4 \beta \Gamma}\,
\exp\left(-\frac{2 q_{max}}{\pi^2\,\beta K}\right),
\label{sigmaRS}
\end{eqnarray}
where $q_{max}=2\pi/a$ is the ultraviolet cut-off in momentum
space. If we now use the earlier estimate for the Frank elastic
constant, $K\,\sim\, k_BT\,Q^2/a$, the solution becomes
\begin{eqnarray}
\sigma\,=\,{4 \beta \Gamma V}\, \exp\left(
-\frac{4}{\pi\,Q^2}\right)\,\propto\,Q^2 \exp[-4/\pi Q^2]
 \,,\label{sigmaRS2}
\end{eqnarray}
where the quadratic order parameter dependance of $\Gamma$ has
been also substituted. This function has a singular dependance on
the order parameter, dropping to zero as $Q\to 0$. As a result,
the disorder is irrelevant close to the nematic-isotropic
transition.

A necessary condition for the RS solution to be applicable is that
the disorder energy must be stable for infinitesimal variations
from that solution. The stability analysis was first introduced in
the context of spin glasses \cite{almeida78} and later applied to
the GVM \cite{mezard91}. First, the variational parameter
$\sigma_{ab}$ is allowed to vary:
$\sigma_{ab}\,=\,\sigma_{ab}^{RS}  \,+\,\epsilon_{ab}$,  with the
replica symmetric ansatz labelled as  $\sigma^{RS}$ and
$\epsilon_{ab}$ the infinitesimal deviation from it.
$F_{var}(\sigma)$ is then expanded to second order in
$\epsilon_{ab}$, naturally, without the linear term, represented
by the zero in the Eq.~(\ref{1diff}):
\begin{eqnarray}
\tilde
F_{var}\,=\,F_{var}(\epsilon=0)\,+\,\frac{1}{2}\,\int_q\sum_{a,b,c,d}
H_{ab\,,\,dc} (q)\ \epsilon_{ab}\,\epsilon_{cd}.
\nonumber
\end{eqnarray}
The  term second-order in $\epsilon_{ab}$ involves the
four-dimensional tensor of coefficients $H_{ab, dc}$, which is
called Hessian and plays a vital role in determining the stability
of the this solution: for RS to be stable, this term must be
positive definite for arbitrary $\epsilon$. This is ensured as
long as all the eigenvalues of the Hessian are non-negative. In
our $XY$ system there are only two non-trivial eigenvalues. One of
them, the so-called replicon eigenvalue, diverges to negative
infinite \cite{mezard91,giamarchi95}:
\begin{eqnarray}
\lambda\,=\,1\,-\,\sigma\,\int
\frac{d^3q}{(2\pi)^2}\,\frac{1}{(\bar{K}q^2)^2} .
\nonumber
\end{eqnarray}
Therefore the replica symmetric solution does not give an
established free energy minimum and is not appropriate.

\subsection{Hierarchical RSB \label{RSBsection}}

The stability analysis of the RS solution, as well as much
previous work on disorder systems, invites us to look for the
hierarchical replica symmetry breaking solutions \cite{parisi80}.
This is known to be the real equilibrium state (i.e. has both
Hessian eigenvalues positive definite) of spin glasses
\cite{dotsenkobook} and random manifolds \cite{carlucci96}. It is
reasonable to expect a similar behavior in our model and therefore
the hierarchical method was chosen to explore replica symmetry
breaking.

In this model the matrix $\sigma_{ab}$ is assumed to have a nested
block-diagonal form with off- and on-diagonal parts map to $
\sigma(v)$ and $\tilde{\sigma}$ respectively, with a continuous
index variable  $v\,\in [0\,,1]$. As a result $G_{ab}^{-1}$ and
its inverse also have this hierarchical form and the quantity of
interest when solving the stationary Eq.~(\ref{sigma}) is now
written as $B_{ab}\,=\,2\, \int_q [\tilde{g}-g(v)]$, where
 \begin{eqnarray}
\fl \tilde{g}\,-\,g(v) =
\frac{1}{\bar{K}\,q^2\,-\,\tilde{\sigma}\,+\,\sigma(1)} +
\int_v^1\,du \, \sigma'(u)\,
\left(\frac{1}{\bar{K}\,q^2\,-\,\tilde{\sigma}\,+\langle
\sigma\rangle\,+\,[\sigma](v)}\right)^2  \label{RSB g-g},
 \end{eqnarray}
where the shorthand notations $\sigma'= d \sigma / dv$ and
$[\sigma](v)\, \equiv \,v\, \sigma(v)\,-\,\int_0^v du\,\sigma(u)$
are used, after Parisi \cite{parisi80}. Finally the condition on
the diagonal part of $\mathbf{\sigma}$ in Eq.~(\ref{sigma_aa})
becomes
\begin{eqnarray}
\tilde{\sigma}\,=\,\int_0^1dv\,\sigma(v)\ =\ \langle \sigma
\rangle . \label{sigma aa RSB}
\end{eqnarray}
It should always be kept in mind that the variable $v$ determines
which diagonal block an element $\sigma_{ab}$ belongs to. The
smaller it is, the closer the element is positioned to the
diagonal. More interestingly, we shall see later that the long
range behavior of the system is specifically associated with these
small values of $v$.

The original stationary equation is written in terms of
hierarchical matrices as:
\begin{eqnarray}
\sigma(v)\,=\,{4 \beta \Gamma V} \,\exp\left\{-4\int_q
\big[\tilde{g}(q)\,-\,g(v)\big]\right\} . \label{generalsigmav}
\end{eqnarray}
Substituting Eqs.~(\ref{RSB g-g}) in (\ref{generalsigmav}):
\begin{eqnarray}
\fl \sigma(v) = {4 \beta \Gamma V}\,\exp\left\{-4\int_q \left[
\frac{1}{\bar{K}\,q^2\,-\,\tilde{\sigma}\,+\,\sigma(1)} + \int_v^1
\frac{\sigma'(u) \,du}{[\bar{K}\,q^2-\tilde{\sigma}+\langle
\sigma\rangle+[\sigma](u)]^2}\right]\right\}. \label{sigmav1}
\end{eqnarray}
Parameter $v$ appears only on the lower limit of the $u$-integral
in Eq.~(\ref{sigmav1}). As a result the stationary equation is
greatly simplified by differentiating it with respect to $v$;
\begin{eqnarray}
\sigma'(v)\,=\,4\,\sigma(v)\int_q\frac{\sigma'(v)}{ [
\bar{K}\,q^2+[\sigma](v)]^2} \ ,\label{bothsolutions}
\end{eqnarray}
where it is noted that Eq.~(\ref{generalsigmav}) shows that
$\sigma(v)$ has no $q$-dependance. This stationary equation for
$\sigma(v)$ has two solutions. The first one is obvious: $ \sigma'
(v)\,=\, 0$. Had this been the real relevant solution, the unique
RS would have been the only possibility (the parameter $\sigma$
does not depend on $v$). However, a second solution also exists
for $\sigma'(v)\neq 0$, and it is given by:
\begin{eqnarray}
1= 4\,\sigma(v)\,\int\frac{dq}{2\pi^2}\,\frac{q^2}
{\left[\bar{K}q^2\,+\, [\sigma](v)\right]^2}\  \Rightarrow \ \
\sigma(v)\,=\,2\pi\,\bar{K}^{3/2}\,[\sigma]^{1/2}
\label{solutionforsigma}
\end{eqnarray}
Differentiating Eq.~(\ref{solutionforsigma}) w.r.t. $v$ and
observing that $[\sigma]'(v)\,=\,v\,\sigma'(v)$, gives the second
solution for $\sigma$:
\begin{eqnarray}
[\sigma](v)\,=\,\pi^2\,\bar{K}^3\,v^2\ \Rightarrow\
\sigma(v)\,=\,2\,\pi^2\,\bar{K}^3\,v\ . \label{second sigma
solution}
\end{eqnarray}
The linear dependance on $v$ is an interesting feature. It implies
that the effect of disorder -- a measure of which is the value of
$\sigma$ -- is smaller for small $v$ (that is, further away from
the diagonal in the discrete form of $\sigma_{ab}$). As we will
see in the next section, the large-distance director correlations
are controlled by the small-$v$ solution. Therefore RSB predicts
that disorder has a milder effect in large-distance correlations,
in marked contrast to the RS solution where $\sigma$ has a
constant value.

\begin{figure}[t]
\begin{center}
\resizebox{0.45\textwidth}{!}{\includegraphics{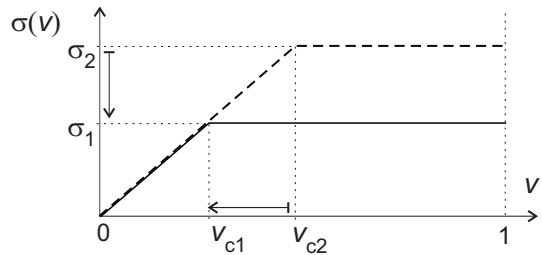}}
 \caption{The
function $\sigma(v)$, plotted for two different disorder strengths
(dashed line for higher $\Gamma$). The weakening disorder
decreases both the crossover point $v_c$ and the plateau value
$\sigma(v_c)$, but leaves the slope of the function unchanged.
}\label{fig sigmaRSB}
\end{center}
\end{figure}

The full RSB form of $\sigma(v)$ incorporates both the linear and
the constant ($\sigma'=0$) solutions. In fact there is a crossover
between the two at a value $v_c$, where $\sigma(v)$ changes from
one regime to the other:
\begin{eqnarray}
\sigma(v)\,=\left\{ \begin{array}{ll}
2\pi\,\bar{K}^3\,v & \mbox{for v $\in [0,v_c]$} \\
2\pi\, \bar{K}^3\,v_c\,=\,const\qquad & \mbox{for v $\in [v_c,1]$}
\end{array}\right. \ ,
 \label{sigmaRSB}
\end{eqnarray}
as sketched in Fig.~\ref{fig sigmaRSB}. This form of branched
function appears in spin glasses \cite{dotsenkobook} as well as
other disorder systems where the GVM has been used, such as random
manifolds \cite{mezard91} and flux lattices \cite{giamarchi95}.

The reason that both solutions are accepted is illustrated in
Fig.~\ref{fig sigmaRSB}. As we shall see shortly, in the
Eq.~(\ref{vc}), $v_c$ reduces to zero in the absence of disorder
$(\Gamma=0)$. This in turn means that $\sigma(v)$ is identically
zero for the whole range of $v\in[0,1]$. The no-disorder limit is
an essential part of an acceptable solution for $\sigma(v)$ and
would be absent if only the linear branch of Eq.~\ref{sigmaRSB}
was accepted.

The only undetermined parameter is the crossover boundary $v_c$,
to which we turn our attention to. Setting $v$ equal to $v_c$ in
Eqs.~(\ref{generalsigmav}) and (\ref{sigmaRSB}) and observing that
they are equal, we find:
\begin{eqnarray}
\sigma(v_c)\,=\,2\,\pi\,\bar{K}^3\,v_c =
\frac{4\,\Gamma}{k_BT}\,\exp \left[-\frac{2 q_{max}}{\pi^2\beta
K}\,+\, \frac{2v_c}{\pi}\tan^{-1}\left(\frac{q_{max}}{\pi \beta K
v_c}\right)\right].
 \nonumber
 \end{eqnarray}
The term linear in the ultraviolet  momentum cutoff
(short-distance, $q_{max}\,=\,2\pi/a$) is identical to
Eq.~(\ref{sigmaRS}) encountered in the Replica symmetry limit.
Since $\beta K\,\sim\, Q^2/a$, this term does not diverge. The
same applies for the argument of the arctangent, which is also
approximated by $2/(Q^2\,v_c)$, after which the equation
determining $v_c$ becomes:
\begin{eqnarray}
v_c \,\simeq\,\frac{2(k_BT)^2\Gamma\,}{\pi K^3}\,\exp
\left[-\frac{4}{\pi\,Q^2}+
\frac{2v_c}{\pi}\tan^{-1}\left(\frac{2}{v_c Q^2}\right)\right].
 \label{vc}
 \end{eqnarray}
The approximately equal sign refers to the use of $K\,\sim\,
Q^2k_BT/a$.

\section{Results and Discussion \label{resultssection}}

\subsection{Director Correlations}

This section discusses how to obtain the domain size as a function
of order parameter and disorder strength. To define the
correlation length we examine how the director correlations decay.
As a result of the use of a single angle to parameterize
$\mathbf{n}$, director correlations are given by
\cite{chudnovsky86,fridrikh97}:
\begin{eqnarray}
C(r) &=& \langle \mathbf{n}(r)\cdot\mathbf{n}(0)\rangle=
\langle\cos\left[\theta(r)-\theta(0)\right]\rangle
\propto e^{-B(r)/2},\nonumber \\
&& {\rm where}\ B(r)\,=\,2\int_q\,G_{aa}\,[1\,-\,\cos(q\,r)]
\nonumber
\end{eqnarray}
In RSB $G_{aa}$ is replaced by (see \cite{mezard91} for detail):
\begin{eqnarray}
\tilde{g}&=& \frac{1}{\bar{K}q^2}\left(1+ \int_0^1\frac{dv}{v^2}
\,\frac{[\sigma](v)}{\bar{K}q^2+
[\sigma](v)}\right)\nonumber \\
&=&\frac{1}{\bar{K}q^2}\bigg[1+\frac{\pi\bar{K}}
{q}\,\tan^{-1}\left(\frac{\pi \bar{K} v_c}{q}\right) +
\left(\frac{1}{v_c}-1\right)
\frac{[\sigma](v_c)}{\bar{K}q^2+[\sigma](v_c)}\bigg].
\label{gtilde}
\end{eqnarray}
The $1/\bar{K}\,q^2$ term which dominates at large $q$ originates
from thermal fluctuations. The remaining terms, which dominate at
small $q$ (large distances), show the effect of disorder.

In order to determine the correlations, $B(r)$ must be calculated
using the random field contribution of Eq.~(\ref{gtilde}). The
$q$-integration is rather involved and yields:
\begin{eqnarray}
B(r) &=& \frac{1}{2}\left\{ \frac{}{} {\rm Euler\Gamma} + SI
\left(\frac{r}{\xi}\right)- CI\left(\frac{r}{\xi}\right) +
\log\left(\frac{r}{\xi}\right)\right. \label{C(r) full}  \\
&-&\left. 2(1-v_c) \sinh\left(\frac{r}{2\xi}\right)\left[
\sinh\left(\frac{r}{2\xi}\right)-
\cosh\left(\frac{r}{2\xi}\right)\right]\right\}, \nonumber\\
 {\rm
where} && \xi^{-1}=\pi K v_c. \nonumber
\end{eqnarray}
Euler$\Gamma \approx 0.58$ refers to the Euler constant, $SI$ and
$CI$ are shorthand notations for the SinhIntegral and CoshIntergal
functions, respectively. Their behavior for small and large
arguments is given by:
\begin{eqnarray}
&&SI(x)\,=\,x\,+\,\mathcal{O}(x^3)\nonumber \\
&&CI(x)\,=\,{\rm Euler}\Gamma \,+\,\log x \,+\,\mathcal{O}(x^2) \
; \ \ {\rm and} \ CI(\infty)\,=\,SI(\infty).\nonumber
\end{eqnarray}

Director correlations decay differently depending how $r$ compares
with the length scale $\xi$, which is identified as the
correlation length or domain size:
\begin{eqnarray}
\left\langle {{\bf{n}}(0) \cdot {\bf{n}}(r)} \right\rangle\,\sim
\, \left\{ {\begin{array}{*{20}c}
   {\exp ( - {\raise0.7ex\hbox{$r$} \!\mathord{\left/
 {\vphantom {r \xi }}\right.\kern-\nulldelimiterspace}
\!\lower0.7ex\hbox{$\xi $}})\ ,   \mbox{for  $r <<\xi$}}\\
   {\left( {  {\raise0.7ex\hbox{$r$} \!\mathord{\left/
 {\vphantom {r \xi }}\right.\kern-\nulldelimiterspace}
\!\lower0.7ex\hbox{$\xi $}}} \right)^{- 1}\ \ \ \, ,   \mbox{for  $r >>\xi$}}\\
\end{array}}  \right.
\label{RSBcorrelations}
\end{eqnarray}
The power-law dependance was first derived by Giamarchi and
LeDoussal \cite{giamarchi95} and is reproduced in our work. It shows
that alignment order persists over greater length scales than one
might first guess from the Larkin argument and the work on random
anisotropy magnets \cite{larkin70,chudnovsky82}. One also associates
the exponential decay at short distances of
Eq.~\ref{RSBcorrelations}, which is also found in the RS case, with
the constant part of $\sigma(v)$ and the algebraic decay with the
linear part. In this context the small-$v$ solution can be taken to
correspond to large distance correlations.

An exact analytic solution of $v_c$ in Eq.~(\ref{vc}) is clearly
impossible, however its dependence on the order parameter $Q$ and
disorder strength $\Gamma$ can be deduced. It is then
straightforward to find $\xi\,=\,\pi K v_c$ as a function of these
two parameters, which is done in the next two sections.

\subsection{Domain size as a function of order parameter}

\begin{figure} 
\begin{center}
\resizebox{0.45\textwidth}{!}{\includegraphics{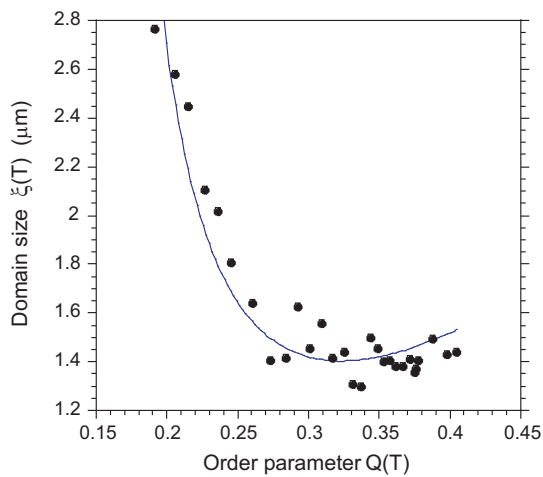}}
 \caption{Combined results of two different data sets: $\xi(T)$
\cite{clarke98} and $Q(T)$  \cite{clarke01} to form a parametric
graph of $\xi(Q)$. The fit (solid line) comes from Eq.~(\ref{xiQ})
and is good for small order parameter where the theoretical
prediction holds.
 \label{fig domain}}
\end{center}
\end{figure}

In this section we present the first theoretical model that
predicts successfully the behavior of the domain size close to the
phase transition.  As discussed in the Introduction, the Imry-Ma
argument gives $\xi \sim K^2/\Gamma \propto Q^2$. This expression
cannot hold close to the nematic-to-isotropic transition since we
expect that $\xi$ should diverge as $Q\to 0$. Experiments
measuring $\xi(Q)$ directly do not exist for nematic elastomers:
it is practically impossible to measure simultaneously the order
parameter and domain size of such systems. However, light
scattering experiments measured $\xi(T)$ the domain size as a
function of temperature \cite{clarke98}. Separately the order
parameter of a \textit{monodomain} elastomer was also measured as
a function of temperature \cite{clarke01}. Assuming that the
polydomain and monodomain samples have the same form of $Q(T)$,
the data are combined with $\xi(T)$ to make a parametric plot of
$\xi(T)$ versus $Q(T)$ in Fig.~\ref{fig domain}. As expected the
domain size increases at small $Q$.

To determine $\xi(Q)$ for small order parameter we must first find
the functional dependence of $v_c$ on $Q$. From their definition
both $v_c$ and $Q$ are smaller than one. Therefore the argument of
the arctangent in Eq.~(\ref{vc}) is large and $\tan^{-1}\to\pi/2$.
Hence
\begin{eqnarray}
v_c \approx \frac{2\Gamma\,}{\pi\,\beta^2 K^3}\,\exp
\left(-\frac{4}{\pi Q^2}+ v_c\right)\ \label{vc small}.
\end{eqnarray}
When $Q\ll 1$ the factor of $\exp(-4/\pi Q^2)$ ensures that $v_c$
is also much smaller than one. Therefore $v_c \ll Q^{-2}$ and term
$v_c$ that sits on the exponent is negligible. Taking into account
that both $\Gamma$ and $K$ are proportional to $Q^2$ for small
$Q$, one finds:
\begin{eqnarray}
\xi\,=\, \frac{1}{\pi K v_c} \,\propto\, Q^2\,e^{\frac{4}{\pi
Q^2}}\label{xiQ}
\end{eqnarray}
As we shall see later, for real elastomers $v_c \approx 10^{-4}$
even for $Q=1$, so taking the arctangent equal to $\pi/2$ is a
very good approximation.

The solid line in Fig.~(\ref{fig domain}) shows a fit to the model
expression (\ref{xiQ}). The agreement between theory and
experiments is good for small $Q$, but breaks down for $Q\geq
0.3$. This is attributed to the fact that $K$ is proportional to
$Q^2$ only for $Q\ll 1$, therefore the pre-exponential factor of
$Q^2$ arising from $K^2/\Gamma$ is strictly valid only for small
$Q$. A similar indirect study of the $\xi(Q)$ was fitted with a
power law $\xi\propto Q^{-2}$ \cite{elias99}. The Imry-Ma
correlation length was considered, forcing the authors to suggest
that $\Gamma \propto Q^6$, which is hard to justify. It is not
reasonable to try to differentiate the two fits, since the data do
not come from direct measurements. However our present model is
theoretically consistent since it assumes $\Gamma\,\propto\, Q^2$
in accordance with the idea of crosslinks providing a random field
that couples to the local order parameter as seen in
Eq.~(\ref{rfenergy}).

\subsection{Domain size as a function of disorder strength}

Lattice Monte Carlo simulations have been performed to study the
equilibrium ordering in a two-dimensional nematic system with
quenched random disorder \cite{yu98}. The long-range correlation
of the director orientation was found to decay as a simple
exponential. However, there is a flaw with the simple exponential
decay, since it does not predict the ``domain size'' to diverge as
$\Gamma\,\to\,0$, as one must expect. The correlation length $\xi$
itself also decays exponentially with increasing strength of the
disordering field. The weak disorder region was never probed
because the amount of computation needed to produce reliable
results increased sharply as $\Gamma$ was reduced.

In order to obtain $\xi$ as a function of $\Gamma$ we must first
find $v_c(\Gamma)$. Going back to Eq.~(\ref{vc}), let us rename
the dimensionless prefactor as
\begin{eqnarray}
D=\frac{2(k_BT)^2\Gamma}{\pi K^3} \label{D}.
\end{eqnarray}
This is a measure of the relative disorder strength. To get an
order of magnitude estimate for $D$ in nematic elastomers, we
consider, following the detailed experimental comparison in
\cite{fridrikh99}, $\gamma\sim 0.4\times k_BT$, $\rho_0\sim
2.5\times 10^{26}m^{-3}$ and $K\sim 10^{-12}Jm^{-1}$, which gives
$D$ of the order of $10^{-4}$. This means that for constant order
parameter $Q=1$ $v_c$ is small and the arctangent in (\ref{vc}) is
equal to $\pi/2$ with good precision.

To obtain a theoretical prediction of $v_c(D)$ let us use a
self-consistent method in Eq.~(\ref{vc}). At $Q=1$ we can write
\begin{eqnarray}
 v_c=D\exp(-4/\pi +v_c) \approx 0.3\,D\,e^{v_c}\label{vc of D1}.
\end{eqnarray}
In the first approximation the right hand-side dependance on $v_c$
is dropped, leaving $v_c\,\approx\,0.3 \,D$. This value of $v_c$
is placed back in the exponent of (\ref{vc of D1}) to give the
estimate
\begin{eqnarray}
v_c\,\simeq\, 0.3 \,D\, e^{0.3 \,D} . \label{vc of D}
\end{eqnarray}
A numerical solution of  Eq.~(\ref{vc}) can be easily found
without the assumption $v_c \ll $1 and the subsequent
approximation of the arctangent term. As seen in Fig.~\ref{fig
vcD}, the agreement of the estimate above with the numerical
solution is very good for small values of $D$, which real systems
are expected to have. An important feature of this plot is that it
starts from the origin. When there is no disorder in the system
($D=0$), $\sigma(v)$ is zero for all $v$ because $v_c =0$.

\begin{figure} 
\begin{center}
\resizebox{0.45\textwidth}{!}{\includegraphics{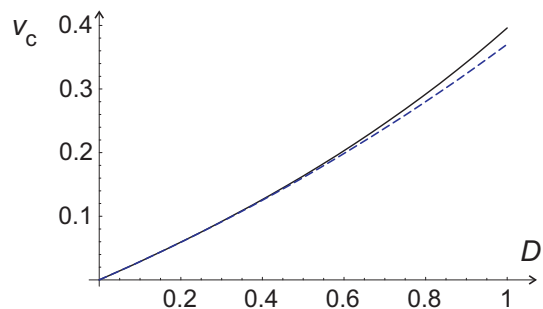}}
\caption{Numerical evaluation of $v_{c}(D)$ from Eq.~(\ref{vc})
for fixed order parameter is plotted as solid line. The dotted
line shows the analytic approximation of Eq.~(\ref{vc of D}). The
agreement between the two curves is so good that one can hardly
distinguish between them in the limit $D \to 0$.
  \label{fig vcD}}
\end{center}
\end{figure}

\begin{figure}  
\begin{center}
\resizebox{0.45\textwidth}{!}{\includegraphics{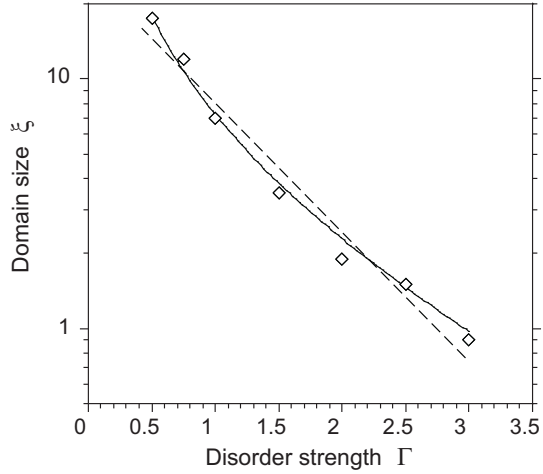}}
 \caption{The log-linear
dependance of the domain size on disorder strength. The data
points are taken from the simulation work \cite{yu98}. The dashed
line is the fit by a simple exponential, provided by the authors
of the original simulation work. The solid line shows the fit with
our model expression (\ref{xi of Gamma}): $ ({9.8}/{\Gamma})
\,e^{-0.45 \Gamma}$ -- a clearly better fit.
  \label{fig xi RSB}}
\end{center}
\end{figure}

The domain size has a more complex functional dependance on
disorder strength than a simple exponential. It is inversely
proportional to $v_c$, therefore since $D \propto \Gamma$:
\begin{eqnarray}
\xi(\Gamma)\propto \frac{K^2}{\Gamma}\,e^{-\alpha\,\Gamma}
\label{xi of Gamma},
\end{eqnarray}
where $\alpha \, \simeq \,0.2\,(\beta^2 K^3)^{-1}$. The
computationally obtained $\xi(\Gamma)$ data points of reference
\cite{yu98} were fitted with the function of Eq.~(\ref{xi of
Gamma}). The resulting fit, Fig.~\ref{fig xi RSB} is an
improvement of the simple exponential function and offers, for the
first time, an analytical explanation for the exponential decay
found by Yu \textit {et al} \cite{yu98}. Importantly the
$(1/\Gamma)e^{-\alpha\Gamma}$ function also predicts the
divergence of the domain size at vanishing disorder, an important
improvement of the previous work.

Nematic elastomers can potentially provide a physical system where
the spin-glass theory can be extensively tested experimentally.
Parameter $\Gamma \sim \gamma^2\rho_0$ can be, in principle,
controlled by changing the density or the molecular nature of
network crosslinks, an important factor in determining many
physical properties of elastomers.

\subsection{Nematic-isotropic phase transition}

\begin{figure}
\begin{center}
\resizebox{0.5\textwidth}{!}{\includegraphics{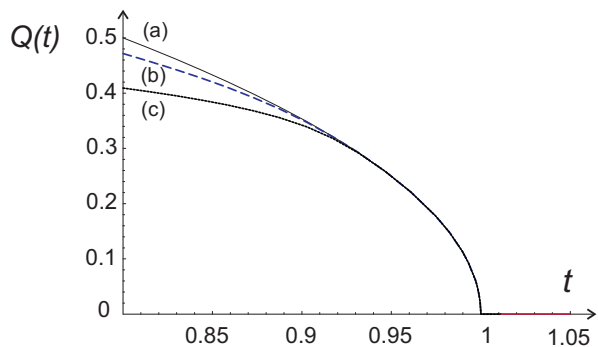}}
\caption{Plots of order parameter against reduced temperature
$t=T/T^*$ that qualitatively show the effect of disorder. Plot
($a$) is obtained from the Landau-DeGennes theory in the absence
of disorder, whereas curves ($b$) and ($c$) show the effect of
increasing disorder.
 \label{fig Q(T)}}
\end{center}
\end{figure}

The last question we address here is whether the presence of
quenched disorder changes the character of the nematic-isotropic
phase transition. In the case of $XY$ nematics whose director is
confined in a plane, the transition is second order because the
cubic term in the Landau-DeGennes expansion of the free energy
density is identically zero: ${\rm tr}
(\underline{\underline{Q}}^3)=0$. We calculate the energy density
arising from disorder as a function of order parameter using the
model Hamiltonian of Eq.~(\ref{G^-1 2d}):
\begin {eqnarray}
\fl \beta F_0 =  \left.\frac{\partial }{\partial
m}\int\mathcal{D}\theta \exp\left(-\frac{1}{2} \sum_{ab}\sum_q
G^{-1}_{ab}\theta_a\theta_b\right) \right|_{m = 0} \propto  \left .
\frac{\partial}{\partial m} \exp\left(-\frac{1}{2}\sum_q tr \log
G^{-1}\right)\,\right|_{m=0}. \label{freeenergy}
\end{eqnarray}
The aim is to add $F_0$ to the Landau-DeGennes expression to see
if there is any significant change on the phase transition. In the
RSB matrix algebra the trace of a logarithm is given by
\cite{mezard91}:
\begin{eqnarray}
 \frac{1}{m}tr\log \mathbf{G}^{-1} = \log\left(\bar{K}q^2
\right)-\frac{\sigma(0)}{\bar{K}q^2}
+\int_0^1\frac{dv}{v^2}\log\left(\frac{\bar{K}q^2}
{\bar{K}q^2+[\sigma](v)}\right).
\end{eqnarray}
Substituting $\sigma(v)$ and integrating over $q$:
\begin{eqnarray}
\fl \frac{1}{m}\int \frac{4\pi q^2\,dq}{(2\pi)^3}\,tr\log
\mathbf{G}^{-1} &=&  \bar
{K}^2\left(\frac{v_c^2}{3}-\frac{v_c}{2}\right) \left[\pi \bar{K}
v_c\tan^{-1} \left(\frac{\pi\bar{
K}v_c}{q_{max}}\right)+q_{max}\right]
\label{F0} \\
&+& \frac{1}{2\pi} \bar{K}q_{max}^2\tan^{-1}\left(\frac{\pi\bar{K}
v_c}{q_{max}}\right) +\frac{q_{max}^3}{6\pi^2}
\log\left(1+\frac{\pi^2\bar{K}^2v_c^2}{q_{max}^2}\right).
\nonumber
\end{eqnarray}
One way to simplify the above expression is to recall that
$v_c\,\beta K/q_{max} \ll  1$ and therefore the arctangent and
logarithmic terms can be expanded in a Taylor series and thus
obtain a $Q$-dependent expression. However, due to the singular
dependence of $v_c$ on the order parameter ($e^{-1/Q^2}$), all the
terms of Eq.~(\ref{F0}) vanish close to the critical point of this
continuous transition. Hence the disorder energy is zero as $Q\to
0$ and the transition remains second order, with unchanged
critical behavior. This is in contrast to the case of
$3D$-director nematics where the addition of disorder changes the
phase transition from first to second order. The overall weak
effect of disorder on the average nematic order in the system is
sketched in Fig.~\ref{fig Q(T)}. Nematic order is weakened away
from the transition, but overall there are no significant changes.

\section{Conclusion \label{sec:7}}

In this paper we have developed a quantitative description for the
ordering of nematics with quenched disorder, following the now
classical work on Replica symmetry breaking in spin glasses. Our
model is focused on a particular case where the director is
confined in a plane (corresponding to the $3D\ XY$ model in
magnets). Such nematic configuration occurs in many experimental
arrangements where thin films are used and is especially relevant
for nematic elastomers. Quenched impurities (chemical crosslinks
in the case of elastomers) act as sources of disorder by providing
easy anisotropy axes. The competition between Frank elasticity and
these random sources leads to the loss of long range orientational
order: the mesogenic units (rod-like molecules) are assumed to be
locally well-ordered with the magnitude of the order parameter $Q$
uniform throughout the system, but the orientation of the director
varies on a length scale $\xi$, called the domain size. We have
extracted the dependence of $\xi$ on $Q$ and $\Gamma$, a measure
of the strength of disorder. In addition we have checked what
effects the addition of disorder has on the nematic-isotropic
phase transition in such $XY$ nematics.

The functional forms of $\xi(Q)$ and $\xi(\Gamma)$ adequately
describe the evolution of the domain size, reported by experiment
and simulations. In particular, $\xi(Q)\sim (K^2/\Gamma) e^{2
q_{max}k_BT/\pi^2 K}\propto Q^2 e^{4/\pi Q^2}$ provides a
consistent account of the divergence of $\xi$ as the phase
transition is approached. For $Q \geq 0.3$, where the exponential
function varies much more smoothly, the Imry-Ma estimate $(\xi_d
\sim K^2/\Gamma)$ is recovered. When disorder is not strong, we
found the domain size to diverge as $\xi(\Gamma)\sim
(K^2/\Gamma)\,e^{-\alpha  \Gamma}$. This form combines the
exponential decay found in some simulations with the classic
Imry-Ma result. It is possible that this behavior can be verified
experimentally by measuring $\xi$ for nematic elastomers with
different crosslink density.

Quenched random-anisotropy effects were found to become irrelevant
as the continuous phase transition of the $XY$ nematic is
approached. The parameter $\sigma$ that models the random field
within the GVM is identically zero as $Q$ approaches zero. For
this reason, the addition of impurities only affects the state of
local order, the $Q(T)$ plot, well below the transition
temperature where $Q\geq 0.3$. In a separate paper, we consider
that nematics with $3D$ director conformation (analogous the the
Heisenberg magnet) and find that they behave in a completely
different way: the presence of disorder transforms the first-order
transition of pure systems to a continuous one.

\section*{Acknowledgements}

This work has been supported by the Leventis foundation, the
Cambridge European Trust and the EPSRC TCM/C3 Portfolio grants. LP
would like to thank Kostas Roussakis, Isaac Perez Castillo and
David Sherrington for many useful discussions.

\section*{References}

\begin{thebibliography}{99}

\bibitem{degen}de Gennes P G and Prost J 1995 {\it The Physics of Liquid Crystals},
(Oxford: Oxford University Press).

\bibitem{kosterlitz}
Kosterlitz J M and Thouless D J 1973 {\em J. Phys. C} {\bf 6}
1181.

\bibitem{kawamura}
Kawamura H and Li M S 2001 {\em Phys. Rev. Lett.} {\bf 87} 187204.

\bibitem{chudnovsky82}
Chudnovsky E M and Serota R A 1982 {\em Phys. Rev. B} {\bf 26}
2697.

\bibitem{larkin70}
Larkin A I 1970 {\em Sov. Phys JETP} {\bf 31} 784.

\bibitem{fridrikh99}
Fridrikh S V and Terentjev E M 1999 {\em Phys. Rev. E} {\bf 60}
1847.

\bibitem{imry75}
Imry Y and Ma S-k 1975 {\em Phys. Rev. Lett.} {\bf 35} 1399.

\bibitem{chudnovsky86}
Chudnovsky E M, Saslow W M and Serota R A 1986 {\em Phys. Rev. B}
{\bf 33} 251.

\bibitem{giamarchi95}
Giamarchi T and Le Doussal P 1995 {\em Phys. Rev. B} {\bf 52}
1242.

\bibitem{feldman00}
Feldman D E 2000 {\em Phys. Rev. B} {\bf 61} 382.

\bibitem{tissier06}
Tissier M and Tarjus G 2006
 {\em Phys. Rev. Lett.} {\bf 96} 087202

\bibitem{uchida00}
Uchida N 2000 {\em Phys. Rev. E} {\bf 62} 5119.

\bibitem{selinger}
Selinger J V, Jeon H G and Ratna B R 2002 {\em Phys. Rev. Lett.}
{\bf 89} 225701.

\bibitem{yu98}
Yu Y K, Taylor P L and Terentjev E M 1998 {\em Phys. Rev. Lett. }
{\bf 81} 128.

\bibitem{disch94}
Disch S, Schmidt C and Finkelmann H 1994 {\em Macromol. Rapid
Comm.} {\bf 15} 303.

\bibitem{lebar05}
Lebar A, Kutnjak Z, Zumer S, Finkelmann H, Sanchez-Ferrer A and
Zalar B 2005  {\em Phys. Rev. Lett.} {\bf 94} 197801.

\bibitem{fridrikh97}
Fridrikh S V and Terentjev E M 1997 {\em Phys. Rev. Lett.} {\bf
79} 4661.

\bibitem{clarke98}
Clarke S M, Terentjev E M, Kundler I and Finkelmann H 1998 {\em
Macromolecules} {\bf 31} 4862.

\bibitem{elias99}
Elias F, Clarke S M, Peck R and Terentjev E M 1999 {\em Europhys.
Lett.} {\bf 47} 442.

\bibitem{03ximu}
Xing X, Mukhopadhyay R, Lubensky T C and Radzihovsky R 2003 {\em
Phys. Rev. E} {\bf 68} 021108.

\bibitem{06biscari}
Biscari P and Terentjev E M 2006 {\em Phys. Rev E} {\bf 73}
120605PRE.

\bibitem{mezard91}
Mezard M and Parisi G 1991 {\em J. Physique I} {\bf 1} 809.

\bibitem{cleaver}
Cleaver D J, Kralj S, Sluckin T J and Allen M P 1997 in: {\em
Liquid Crystals in Complex Geometries Formed by Polymer and Porous
Networks}  (Taylor \& Francis).

\bibitem{edwards88}
Edwards S F and Muthukumar 1988 {\em J. Chem Phys.} {\bf 89} 2435.

\bibitem{edwards75}
Edwards S F and Anderson P W 1975 {\em J. Phys. F} {\bf 5} 965.

\bibitem{almeida78}
de Almeida J R L and Thouless D J 1978 {\em J. Phys. A} {\bf 11}
983.

\bibitem{parisi80}
Parisi G 1980 {J. Phys. A} {\bf 13} 1101; {\it ibid}: L115.

\bibitem{dotsenkobook}
Dotsenko V 2001 {\em Introduction to the Replica Theory of
Disorderred Statistical Systems} (Cambridge: Cambridge University
Press).

\bibitem{carlucci96}
Carlucci D M, de Dominicis C and Temesvari T 1996 {\em J. Physique
I} {\bf 6} 1031.

\bibitem{clarke01}
Clarke S M, Hotta A, Tajbakhsh A R and Terentjev E M 2001 {\em
Phys. Rev. E} {\bf 64} 061702.

\end{thebibliography}

\end{document}